\documentclass[prints]{aa}
\usepackage{psfig}
\usepackage{graphicx}
\usepackage{supertabular}

\begin{document}

\title{The Bulk Motion of Flat Edge-On Galaxies Based
on 2MASS Photometry}
\titlerunning{The Bulk Motion of Flat Edge-On Galaxies}

\author{Yu.N. Kudrya \inst{1}
\and V.E. Karachentseva \inst{1}
\and I.D. Karachentsev \inst{2}
\and S.N. Mitronova \inst{2,5}
\and T.H. Jarrett \inst{3}
\and W.K. Huchtmeier \inst{4}}
\authorrunning{Yu. N. Kudrya et al.}
\institute{Astronomical Observatory of Kiev Taras Shevchenko
National University, 04053, Observatorna str., 3, Kiev, Ukraine
\and Special Astrophysical Observatory, Russian Academy of
Sciences, N.Arkhyz, KChR, 369167, Russia
\and Infrared Processing
and Analysis Center, Mail Stop 100-22, California Institute of
Technology, Jet Propulsion Laboratory, Pasadena, CA 91125
\and Max-Planck - Institut f\"{u}r Radioastronomie, Auf dem H\"{u}gel
69, 53121, Bonn, Germany
\and Isaac Newton Institute of Chile, SAO Branch, Russia}
\date{submitted, accepted}
\abstract{ We report the results of applying the 2MASS
Tully-Fisher (TF) relations to study the galaxy bulk flows. For
1141 all-sky distributed flat RFGC galaxies we construct $J, H,
K_s$ TF  relations and find that Kron $J_{fe}$ magnitudes show the
smallest dispersion on the TF diagram. For the sample of 971 RFGC
galaxies with $V_{3K} <$ 18000 km s$^{-1}$ we find a dispersion
$\sigma_{TF}=0.42^m$ and an amplitude of bulk flow $V$= 199$\pm$61
km s$^{-1}$, directed towards $l=301\degr\pm18\degr,
b=-2\degr\pm15\degr$. Our determination of low-amplitude coherent
flow is in good agreement with a set of recent data derived from
EFAR, PSCz, SCI/SCII samples. The resultant two- dimensional
smoothed peculiar velocity field traces well the large-scale
density variations in the galaxy distributions. The regions of
large positive peculiar velocities lie in the direction of the
Great Attractor and Shapley concentration. A significant negative
peculiar velocity is seen in the direction of Bootes and in the
direction of the Local void. A small positive peculiar velocity
(100 -- 150 km s$^{-1}$) is seen towards the
 Pisces-Perseus supercluster, as well as the Hercules - Coma - Corona
  Borealis supercluster regions.
}
\maketitle

\section{Introduction}

Since the pioneering work by Aaronson, Mould, Huchra and
collaborators (1979, 1980, 1982), the infrared Tully-Fisher
relation (IRTF) has been widely used in the study of galactic bulk
flows on different scales. A rather complete review on the IRTF
application is given in the thesis of Bamford (2002).

The cosmic flow investigations require both enormous and
homogeneous samples. One way to increase the number of
observational data is the sample combination procedure. For
example, making the Mark III catalogue, Willick et al. (1997)
compiled the infrared magnitudes and line widths from various
sources and reduced them to a uniform system.

The appearance of the complete and homogeneous 2MASS survey Skrutskie
et al. 1997
opens up new opportunities for cosmic flow study by IRTF. In our
previous work (Karachentsev et al., 2002, hereafter Paper I) we
identified the spiral edge-on galaxies from ``The Revised Flat
Galaxy Catalogue'', RFGC (Karachentsev et al., 1999) with the
objects from ``~The Extended Source Catalog'', XSC (Cutri et al.,
1998; Jarrett et al., 2000). Of the total
number of 4236 RFGC galaxies, 2996  galaxies (e. g. 71\%) have
been detected in the $J, H, K_{s}$-bands. The RFGC catalogue was
created using the material of the photographic sky surveys POSS-I
and ESO/SERC. In Paper I we analyzed in detail the
2MASS-characteristics of flat galaxies entered in the XSC. To
build the $B$-, $I$-, $J$-, $H$-, $K_{s}$- TF relations , we used:
$B_t$- magnitudes from RFGC, calculated from angular diameters,
taking into account the galaxy surface brightness, as well as
type, and other galaxy parameters; the total $I$-magnitudes from
Mathewson {\&} Ford (1996) and Haynes et al. (1999); the isophotal
$J$-, $H$-, $K_{s}$- magnitudes measured in elliptical apertures
at a level of $K_{s}$=20 mag/arcsec$^{2}$. For a sample of 436
flat galaxies with this set of magnitudes and known radial
velocities and HI line widths we obtained the slope of the linear
TF regression, increasing from 4.9 in the $B$-band to 9.3 in the
$K_{s}$- band. The derived scatter on the TF diagram did not show
a significant decrease from the blue to infrared band, and after
excluding of dwarf galaxies reached $\sim $ 0.6 mag.

In the present work we study in dipole approximation the bulk
motion of flat galaxies from a homogeneous sample -- the RFGC
catalogue using the $J$-, $H$-, $K_{s}$-TF relations based on the
2MASS photometry data. We show that the scatter on the TF diagram
can be diminished significantly by two factors: a cleaning of the
initial sample and including additional photometric parameters in
the simple TF relation. As a result, for 971 all-sky distributed
flat galaxies the dipole solution is:  $V$=199 km s$^{-1}$,
$l=301\degr$, $b= -2\degr$. The smoothed peculiar velocity field
repeats, as a whole, the large-scale distribution of flattened
galaxies from the 2MASS.

\section{The sample}

\subsection{The HI line widths (the sample RFGC-W$_{50})$.}

In Paper I we presented the radial velocities of 1772 RFGC galaxies.
However, not all of these 1772
objects have the HI line width measurements. To compile the list
of the RFGC galaxies with known estimates of $W_{50}$ (measured
directly or calculated from rotation curves), we use the following
sources:

 a) The list of flat galaxies with known velocities and line widths
(Karachentsev et al., 2000a). This compilation consists of several subsamples
observed by Giovanelli et al., 1997a, Makarov et al., 1997\,a,\,b; 1999, 2001,
Mathewson et al. (1992), Mathewson and Ford (1996), and Matthews and van
Driel (2000).
 b) Our identifications of southern RFGC galaxies with the HIPASS survey
sources (Karachentsev {\&} Smirnova, 2002), and with ``A Catalog of
HI-Selected Galaxies from the South Celestial Cap Region of Sky''
(Kilborn et al., 2002).
 c) The last version of LEDA database (Paturel et al. 1996).
 d) Unpublished data on the HI observations of RFGC galaxies at the Effelsberg
and Nan\c{c}ay radio telescopes (Huchtmeier et al., 2003).
The sources a),b),c), and d) contain, respectively, 78\%, 2\%, 11\%, and 9\%
of the whole sample.
A total of 1653 sets of radial velocities, $V_{h,}$ and line
widths, $W_{50, }$ were used in the initial list, including some
multiple observations of one and the same galaxy. To check the
``best'' estimate among double and triple ones, we determine in a
dipole approximation the distances $Hr$, using the TF relation
``linear diameter - line width'' from Karachentsev et al. (2000b).
The observed radial velocities were reduced to the cosmic
microwave background $3K$ system according to Kogut et al. (1993),
and the observed line widths were corrected for cosmological
broadening and turbulence following Tully \& Fouqu\'e (1985). No
corrections for inclination were made because the RFGC galaxies
with their apparent optical axis ratio $a/b \geq 7$ are by
definition very much inclined to the line of sight ($i\geq
82\degr$). The galaxy peculiar velocities were calculated as
$V_{pec}=V_{3K}-Hr$. Here and hereafter the inferred distance $Hr$
is expressed in km s$^{-1}$. We retained in the RFGC-W$_{50}$
sample only the galaxies from multiple observations whose peculiar
velocities were minimal. In all the cases we took into account the
galaxy morphology, a possible confusion from a near neighbour, the
signal/noise ratio etc. After excluding multiple measurements,
1617 RFGC galaxies with known radial velocities and line widths
were entered in the RFGC-W$_{50}$ sample.

\subsection{The 2MASS photometry ( the sample RFGC-2MASS)}

To compile this sample, we performed the cross-identification
between RFGC and XSC catalogues. The initial file contains 3001
lines with 2MASS data. We used for the processing the following
characteristics (Jarrett et al., 2000, 2003):

$r_{20}$ -- major isophotal radius in arcsec, neasured at the
20 mag/arcsec$^{2}$ level in the $K_s$ band via photometry in elliptical
isophotes;

$r_{fe}$ -- fiducial elliptical Kron radius in arcseconds;

$r_{ext}$ -- radius of the ``total'' aperture in arcseconds;

$J_{20fe}$, $H_{20fe}$, $K_{20fe}$ -- isophotal fiducial
elliptical-aperture magnitudes in corresponding bands measured at
the $K_{s}$- band fiducial 20 mag/arcsec$^{2}$ isophotal radius;

$J_{fe}$, $H_{fe}$, $K_{fe}$ -- Kron fiducial elliptical-aperture
magnitudes measured at the $K_{s}$- band fiducial elliptical Kron
radius;

$J_{ext}, H_{ext}, K_{ext}$  --- integral "total" magnitudes as derived
        from the isophotal magnitudes ($J_{20}, H_{20}, K_{20}$) and
        the extrapolation of the fit to the radial surface brightness
        distribution. The extrapolation ($r_{ext}$) is carried out to
        roughly four times the disk scale length. (Details are given in
	Jarrett et al. 2003);

\textit{Jhl} -- $J$-band ``effective'' surface brightness (at $J$-
band half-light ``effective'' radius);

\textit{Jcdex} -- $J$-band concentration index (3/4 vs. 1/4 light
radius);

\textit{sba} -- axis ratio ($b$/$a)$ for the $J+H+K_{s}$ combined
image (``super'' coadd).

A comparison of two lists, RFGC-2MASS and RFGC-W$_{50 ,}$ gives 1215 common
galaxies, including 68 galaxies with multiple 2MASS estimates. The selection
of the best data among the duplicate ones is made by comparing the
deviations from the simple linear TF relation:

$$M=C_{1}+C_{2} logW^{c} , \eqno(1)$$ where $$M=m-A-
25-5log(V_{3K}/H_{o}), \eqno(2)$$

$A$ is extinction and $H_0$=75 km s$^{ - 1}$Mpc$^{ - 1}$.
It should be noted that the equation (2) is known to be incorrect
at small distances since nearby galaxies around the Local Group do not
need to have an additional $\sim$ 600 km s$^{-1}$ correction subtracted
from their apparent velocities to be placed in the proper Hubble flow.
Altogether, 1141 galaxies were entered in the joint
RFGC-W$_{50}$-2MASS sample after the excluding procedure. Some
parameters of different observables are given in
Table 1, where $\sigma$ is the standard deviation.

\begin{table}[htbp]
\centering
\caption{Statistical characteristics of the flat galaxy sample}
\begin{tabular}
{|p{70pt}|p{25pt}|p{25pt}|p{25pt}|p{25pt}|} \hline N=1141& min&
max& mean& $\sigma$ \\ \hline $J_{20fe}$, mag& 7.5& 15.9& 13.1&
1.21 \\ \hline $H_{20fe}$, mag& 6.6& 15.4& 12.3& 1.23 \\ \hline
$K_{20fe}$, mag& 6.3& 14.9& 12.0& 1.23 \\ \hline $J_{fe}$, mag&
7.4& 15.2& 12.9& 1.10 \\ \hline $H_{fe}$, mag& 6.6& 14.7& 12.1&
1.13 \\ \hline $K_{fe}$, mag& 6.3& 14.4& 11.8& 1.13 \\
 \hline $J_{ext}$, mag& 7.4& 15.9& 12.9& 1.13 \\
 \hline $H_{ext}$, mag& 6.5& 15.3& 12.1& 1.16 \\
 \hline $K_{ext}$, mag& 6.2& 14.8& 11.8& 1.18 \\
 \hline lg($r_{ext}$, arcsec)& 0.82& 2.36& 1.57& 0.22 \\
 \hline lg($r_{fe}$, arcsec)& 0.81& 2.08& 1.50& 0.21 \\
 \hline lg($r_{20fe}$, arcsec)& 0.70& 2.04& 1.36& 0.25 \\
 \hline
\textit{Jhl}, mag/arcsec$^{2}$& 16.8& 20.6& 19.0& 0.68 \\ \hline
\textit{Jcdex}& 1.89& 8.69& 3.99& 0.92 \\ \hline \textit{sba}&
0.10& 1.00& 0.24& 0.097 \\ \hline lg($W_{50}$, km s$^{ - 1})$&
1.36& 2.89& 2.50& 0.163 \\ \hline $W_{50}$, km s$^{ - 1}$& 23&
782& 335& 112 \\ \hline lg($V_{3K}$, km s$^{ - 1})$& 2.24& 4.38&
3.73& 0.269 \\ \hline $V_{3K}$, km s$^{ - 1}$& 175& 23758& 6341&
3327 \\ \hline
\end{tabular}
\label{tab1}
\end{table}
As known, the total and the Kron magnitudes are something like
15 -- 20 \% brighter than the isophotal magnitudes (see figures 11, 12 and 13
in http://spider.ipac.caltech.edu/staff/jarrett/papers/LGA
/LGA\_fig.htm).

\section{Construction of the optimal sample and calculation of the dipole
parameters}

For each galaxy of our sample we have a set of the isophotal, Kron, and
the total magnitudes in the $J, H, K_s$- bands. The check of mutual
correlations between all the magnitudes shows that they are well
correlated (Table 2).

\begin{table*}
\centering
\caption{Some parameters of mutually correlated linear dependence $y =
kx + c$ between different magnitudes  for the sample of 1141 galaxies}
\begin{tabular}
{|l|l|l|l|l|l|l|l|} \hline \multicolumn{1}{|c|}{$x$}&
\multicolumn{1}{|c|}{$y$}& \multicolumn{1}{|c|}{$\sigma_{fwd}$}&
\multicolumn{1}{|c|}{$\sigma_{inv}$}&
\multicolumn{1}{|c|}{$\sigma_{orth}$}& \multicolumn{1}{|c|}{$r$}&
\multicolumn{1}{|c|}{$k$}& \multicolumn{1}{|c|}{$c$} \\ \hline
$J_{20fe}$& $H_{20fe}$& 0.119& 0.117& 0.083& 0.9954&
1.019$\pm$0.003& $-$1.042$\pm$0.039\\ $J_{20fe}$& $K_{20fe}$&
0.162& 0.159& 0.114& 0.9913& 1.019$\pm$0.004& $-1.404\pm$0.053\\
$H_{20fe}$& $K_{20fe}$& 0.114& 0.114& 0.080& 0.9958&
1.000$\pm$0.003& $-0.361\pm$0.034\\ \hline
 $J_{fe}$& $H_{fe}$& 0.131&
0.127& 0.091& 0.9933& 1.026$\pm$0.004& $-1.107\pm$0.046\\
$J_{fe}$& $K_{fe}$& 0.173& 0.168& 0.121& 0.9882& 1.026$\pm$0.005&
$-1.460\pm$0.061\\
 $H_{fe}$& $K_{fe}$& 0.131& 0.131& 0.093&
0.9932& 1.000$\pm$0.003& $-0.352\pm$0.042\\ \hline
 $J_{ext}$& $H_{ext}$& 0.197&
0.193& 0.138& 0.9854& 1.025$\pm$0.005& $-1.091\pm$0.068\\
$J_{ext}$& $K_{ext}$& 0.223& 0.214& 0.155& 0.9820&
1.043$\pm$0.006& $-1.648\pm$0.077\\
 $H_{ext}$& $K_{ext}$& 0.190& 0.187& 0.133&
0.9870& 1.018$\pm$0.005& $-0.535\pm$0.060\\  \hline
 $J_{20fe}$& $J_{fe}$&
0.146& 0.161& 0.108& 0.9912& 0.905$\pm$0.004& $-1.029\pm$0.047\\
$H_{20fe}$& $H_{fe}$& 0.151& 0.165& 0.111& 0.9910&
0.911$\pm$0.004& +0.898$\pm$0.045\\
 $K_{20fe}$& $K_{fe}$& 0.143&
0.157& 0.106& 0.9919& 0.911$\pm$0.003& +0.874$\pm$0.042\\ \hline
 $J_{fe}$& $J_{ext}$& 0.165&
0.160& 0.115& 0.9894& 1.030$\pm$0.004& -0.414$\pm$0.058\\
 $H_{fe}$& $H_{ext}$& 0.177&
0.172& 0.124& 0.9883& 1.030$\pm$0.005& -0.373$\pm$0.057\\
 $K_{fe}$& $K_{ext}$& 0.150&
0.144& 0.104& 0.9919& 1.048$\pm$0.004& -0.544$\pm$0.047\\
 \hline
\multicolumn{8}{|l|}{{\em Notes to table:} $\sigma_{fwd}$, $\sigma_{inv}$, $\sigma_{orth}$
are the standard deviations (in mag) from the direct,}\\
\multicolumn{8}{|l|}{inverse and orthogonal TF regressions, respectively;
$r$ is the coefficient of correlation;}\\
\multicolumn{8}{|l|}{$k$ and $c$ are the coefficients of the orthogonal regression.}\\
\hline
\end{tabular}
\end{table*}

As an illustration of tight mutual relationship between the 2MASS
magnitudes, in Fig.1 we give the regression of $K_{20fe}$ on
$H_{20fe}$, consist with the NIR colors of disk galaxies (Jarrett 2000;
see also Fig. 20 in Jarrett et al. 2003).

\begin{figure}
\includegraphics[width=8.5cm]{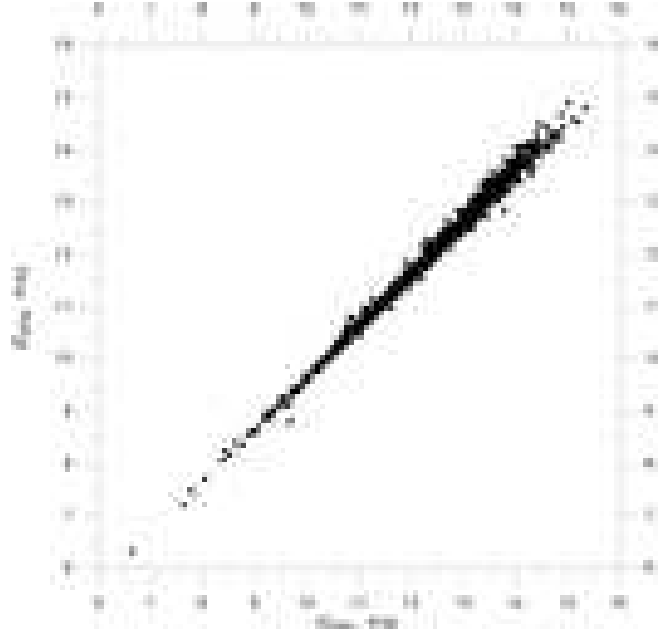}
\vspace{5mm}
\caption{ The relationship between $H_{20fe}$ and $K_{20fe}$.}
\end{figure}

According to the data in Table 2, all three color bands seem to be of
equal value.

To select between the nine different kinds of magnitudes, we built TF
relation (1) for each of them and calculated the characteristics
of bulk motion in a dipole approximation. At this stage the
apparent magnitudes were corrected for Galactic extinction as

$$J^c=J - 0.207\cdot A_B,$$ $$ H^c=H - 0.132\cdot A_B,\eqno(3)$$
$$K^c=K - 0.084\cdot A_B,$$
where $A_B$ is an extinction in the $B$-band according to
Schlegel et al. (1988). Here we did not correct
the magnitudes for internal galaxy extinction, because the internal
extinction depends on the galaxy inclination, as well as the galaxy
luminosity (Verheijen, 2001), which is a priori unknown.

For each of magnitudes we use  constructed TF relations to calculate the
characteristics of bulk motion in a dipole approximation. We adopt the
simple linear model: $V_{pec}=\vec V \vec e_r+\delta$, where
$\vec V$ - the bulk velocity to the apex, $e_r$ - an unit radial
vector of galaxy direction. The value of $\Sigma \delta^2$ with
summing on the whole sample have been minimized by a least squares
method.

The results of calculations are given in Table 3.
\begin{table*}
\centering \caption{Parameters of the TF relation (1) and the
dipole characteristics for 1141 flat galaxies.}
\begin{tabular}{|l|l|l|l|l|l|l|l|} \hline
&\multicolumn{1}{|c|}{$\sigma_{TF}$}& \multicolumn{1}{|c|}{$C_2$}&
\multicolumn{1}{|c|}{$\sigma_{V}$}& \multicolumn{1}{|c|}{$V$}&
\multicolumn{1}{|c|}{$l$}& \multicolumn{1}{|c|}{$b$}&
\multicolumn{1}{|c|}{$F$}\\ \hline $J_{fe}$& 0.855& $-6.24\pm$0.15
(1653)& 1940& 416$\pm$106& 284$\pm$15& $-21\pm$12& 5.2\\
$J_{20fe}$& 0.940& $-6.81\pm$0.17 (1630)& 2177& 568$\pm$118&
292$\pm$12& $-24\pm$10& 7.8\\$J_{ext}$& 0.911& $-6.50\pm$0.16
(1581)& 2076& 437$\pm$113& 287$\pm$15& $-17\pm$12& 5.0\\\hline
 $H_{fe}$& 0.887& $-6.50\pm$0.16
(1668)& 2033& 448$\pm$111& 287$\pm$15& $-23\pm$12& 5.5\\
$H_{20fe}$& 0.972& $-7.05\pm$0.17 (1635)& 2267& 565$\pm$122&
293$\pm$13& $-24\pm$10& 7.2\\ $H_{ext}$& 0.947& $-6.78\pm$0.17
(1592)& 2183& 443$\pm$119& 288$\pm$16& $-24\pm$13& 4.7\\\hline
 $K_{fe}$& 0.907& $-6.68\pm$0.16
(1686)& 2062& 340$\pm$113& 285$\pm$19& $-22\pm$15& 3.1\\
$K_{20fe}$& 0.990& $-7.21\pm$0.18 (1647)& 2295& 475$\pm$124&
292$\pm$15& $-23\pm$12& 4.9\\ $K_{ext}$& 0.966& $-6.99\pm$0.17
(1625)& 2226& 378$\pm$121& 290$\pm$19& $-22\pm$15& 3.3\\
 \hline
\end{tabular}
\end{table*}

The columns of Table 3 denote:

(1) -- 2MASS magnitude, by which the absolute magnitude in the TF
relation was calculated;

(2) -- dispersion on the TF diagram in mag;

(3) -- slope C$_{2}$ in formula (1) and its statistical
significance according to the Fisher criterion in parentheses;

(4) -- dispersion of peculiar velocities in km s$^{-1}$, which
includes the error in the measured distances;

(5) -- modulus of the bulk motion velocity and its error in km
s$^{-1}$;

(6), (7) -- galactic longitude and latitude of the apex and their
errors in degrees;

(8) -- significance of the vectorial dipole solution according to
the Fisher criterion. (Note that for confidence probability
95{\%},  the quantile of the Fisher distribution  is equal to 2.6
for three degrees of freedom for numerator and infinity ones for
denominator).

The errors in $V$, $l$, and $b$ were calculated by first calculating
the diagonal elements $B_{VV}, B_{ll}$ and $B_{bb}$
of covariance matrix $\bf B$ in the frame \{$\vec{e}_V, \vec{e}_l,
\vec{e}_b$\},and then $\Delta V = (B_{VV})^{1/2}, \Delta l =
\arctan\{ (B_{ll})^{1/2}/
  V\}, \Delta l = \arctan \{(B_{bb} )^{1 / 2} / V\}. $

As seen from Table 3, for every photometric magnitude,
the dispersion in the TF fit is too large, nearly
1.5 times as high as that presented in Paper I. We suggest therefore
that the dipole parameters in Table 3 be considered only preliminary.
Due to a larger dispersion, the slopes of the TF relation are
flatter than those obtained in Paper I$. $

It is also seen that the Kron magnitudes (especially $J_{fe})$
have a slightly lower dispersion in the TF diagram than isophotal
as well as extended magnitudes. We built the diagrams of the
residuals ``isophotal minus Kron magnitude'' depending on the
absolute isophotal magnitude. For all three color bands in a wide
range of absolute magnitudes {\{}-16$^{m}$ , -20$^{m}${\}} the
Kron magnitudes are on average $\sim $0.4$^{m}$ brighter, thus
recovering a significant fraction of the disk light that is lost
in the background noise (see also Fig.12 in Jarrett et al., 2003)
. This can reduce partially the known non-linearity of the TF
relation at the lower luminous end.

We adopt $J_{fe}$ as the basic apparent magnitude and accomplish
the procedure of trimming of combined RFGC-W$_{50}$-2MASS sample.
The galaxies were omitted from the sample by such criteria:

1. $W_{50} \leq$ 100 km s$^{-1}$ (N=13). These are bluish dwarf
galaxies with large uncertainties in both the profile width and
the IR magnitudes.

2. The deviation from the TF relation is more than 3$\sigma$
(N=63).

3. $\vert V_{pec}\vert \geq$ 3000 km s$^{-1}$ (N=119). We assume
that such large peculiar velocity values are due to non-physical
reasons.

We also eliminate 6 galaxies with $V_{3K} \geq 18000$ km s$^{-1}$
and (symmetrically) 10 galaxies with $Hr\geq$ 18000 km s$^{-1}$ to
diminish the incompleteness of the sample at large distances.
The process of elimination was converged after 5 steps;
altogether, taking into account the cross-sections, we excluded 170
galaxies.

Excluding the dwarf galaxies with smallest  velocity widths means
simultaneously eliminating the nearest galaxies. For new sample of
971 galaxies the minimal value of $V_{3k}$ increases to 494 km s$^{-1}$ in
comparison with 175 km s$^{-1}$ for non-cleaned sample. Note that
from the total number of 971 galaxies, about 87\% galaxies have
$V_{3K}> 3000$ km s$^{-1}$, i.e. lie outside the Local
supercluster. Thus, we believe the contribution of nearby galaxies in
the bulk motion parameters (when the CMB frame is used) is not
significant.

\begin{figure}[htbp]
\includegraphics[width=8.5cm]{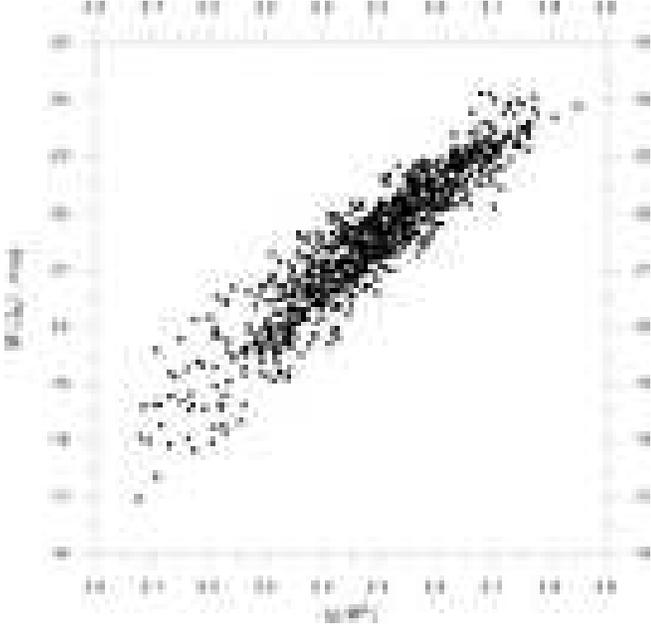}
\vspace{5mm}
\caption{ The Tully-Fisher relation for 971 flat edge-on galaxies.}
\end{figure}

The TF relation for a new sample of  971 galaxies as $M(J_{fe})$
vs. log $W^{c}$ is given in Fig.~2. It has a slope $C_{2} = -8.1$
and a dispersion $\sigma_{TF}$=0.430. The bulk motion parameters
obtained from this relation are:
 $$\sigma_{V}= 1062 km \cdot s^{-1}, V= 323 km \cdot s^{-1},$$
 $$l = 293\degr, b = -15\degr.$$

As is seen in Fig.2, eliminating 15\% of the initial sample using
the "clean" criteria given above, greatly reduces the dispersion
in the TF relation.

Next, for the sample of N=971 we consider seven-parametric
generalized TF relation, in which photometric characteristics and
color index are entered:

$$M(J_{fe})=C_1+C_{2}\cdot lg(W^c)+C_3\cdot jhl+C_4\cdot jcdex+$$
$$+C_{5}\cdot (J_{fe}-K_{fe})+C_6\cdot sba+C_{7} \cdot lg(r_{ext})
\eqno(4)$$.

For the seven- parametric TF relation (4) the solutions are: the
slope $C_2= -7.6, \sigma_{TF}=0.423,$ $\sigma_{V}$= 1044 km
 s$^{-1}$, $ V $ = 187 km  s$^{-1}$, $l = 300\degr, b =
-1\degr$. The $Jhl$, ($J_{fe}-K_{fe})$ and $sba$ terms are
statistically insignificant. In this case, only four terms remain
in the new generalized TF relation:

$$M(J_{fe})=C_{1}+C_{2}\cdot\lg(W^{c})+C_{3}\cdot
jcdex+C_{4}\cdot\lg(r_{ext}). \eqno(5)$$

The new parameters are practically the same as for (4): zero-point
$C_{ 1}= -1.6$, slope $C_{ 2}= -7.6$, $C_{ 3}=-0.046$,
$C_4=-0.44$, $\sigma_{TF}$= 0.422, $\sigma_{V}$=1045 km s$^{-1}$,
$V $ = 199 km s$^{-1}$, $l = 301\degr$, $b = -2\degr$.

The significance of the $jcdex$ term ($F$=4.5) means the TF
relations depend on the Hubble type, namely the central bulge to disk
distribution. Statistical insignificance of the $sba$ term
($F$=0.3) indicates that internal extinction in the infrared does not
correlate, surprisingly, with the infrared axial ratio. The influence
of the surface brightness term, $jhl$, in the regression (4) falls, but
its role is partially evident via the lg($r_{ext})$ term
($F$=17.5).
In Fig.3 we give distances $Hr$ derived from TF relation (5)
depending on radial velocities $V_{3K}$ for the sample N=971.

\begin{figure}[htbp]
\includegraphics[width=8.5cm]{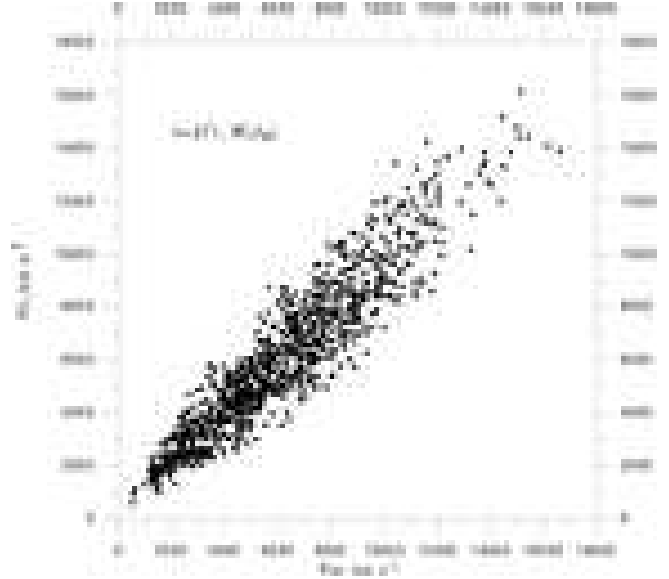}
\vspace{5mm}
\caption{
The relationship between the derived
distances $Hr$ and radial velocities $V_{3K}$.}
\end{figure}

We also built analogous (seven- and four- parametric) TF relations
based on the other eight 2MASS photometric measures. They
demonstrate the same tendencies: the terms $Jhl$,
($J_{fe}-K_{fe})$, $sba$ also are insignificant and the dipole
bulk motion parameters are near to ones mentioned above.

The results of calculations with nine TF relations (5) for the
sample of galaxies with $V_{3K }<$ 18000 km s$^{ - 1}$ and $Hr <$
18000 km s$^{ - 1}$ are collected in Table 4. The designations are
the same as in Table 3.

\begin{table*}
\centering \caption{Parameters of TF regressions (5) for the 2MASS
magnitudes and the dipole characteristics for 971 flat galaxies.}
\begin{tabular}{|l|l|l|l|l|l|l|l|} \hline
& \multicolumn{1}{|c|}{$\sigma_{TF}$}&
\multicolumn{1}{|c|}{$C_2$}& \multicolumn{1}{|c|}{$\sigma_{V}$}&
\multicolumn{1}{|c|}{$V$}& \multicolumn{1}{|c|}{$l$}&
\multicolumn{1}{|c|}{$b$}& \multicolumn{1}{|c|}{$F$} \\ \hline
$J_{fe}$& 0.422& $-7.64\pm0.13$ (3558)& 1045 & 199$\pm$61&
$301\pm$18 & $-2\pm$15& 3.5\\ $J_{20fe}$& 0.457& $-8.21\pm$0.14
(3505)& 1144& $248\pm67$& $306\pm16$& $-5\pm13$& 4.6\\ $J_{ext}$&
0.464& $-7.85\pm$0.14 (3111)& 1153& $215\pm67$& $302\pm18$&
$+7\pm15$& 3.5\\ \hline $H_{fe}$& 0.437& $-7.86\pm$0.13 (3509)&
1078 & 209$\pm$64& 294$\pm$18& $-9\pm16$& 3.6\\ $H_{20fe}$& 0.471&
$-8.45\pm$0.14 (3490)& 1167 & 239$\pm$69& $304\pm$17& $-7\pm$16&
4.0\\ $H_{ext}$& 0.485& $-8.15\pm$0.15 (3067)& 1194 & 198$\pm$70&
$288\pm$20&  $0\pm$17& 2.7\\ \hline $K_{fe}$& 0.445&
$-8.14\pm$0.13 (3635)& 1116& $153\pm66$& 294$\pm$24& $-1\pm$20&
1.8 \\ $K_{20fe}$& 0.483 & $-8.72\pm0.15$ (3552)& 1209 &
201$\pm$71& $300\pm$20& $-3\pm$18& 2.7\\ $K_{ext}$& 0.480 &
$-8.34\pm0.15$ (3289)& 1204 & 138$\pm$69& $301\pm$28& $+12\pm$24&
1.4\\ \hline
\end{tabular}
\end{table*}

We constructed also the second sample with the same restrictions
as for the sample $N$=971, but with $V_{3K }<$ 12000 km s$^{-1}$
and $Hr<$12000 km s$^{-1}$, to diminish the influence of data
incompleteness at large distances. In Table 5 we show
the same data as in Table 4 but for the sample of galaxies
with $V_{3K }<$ 12000 km s$^{ - 1}$ and $Hr<$12000 km
s$^{ - 1}$ .The designations are the same as in Table~3. The
elimination of 50 remote galaxies does not change the dipole
parameters much as well as the TF parameters $C_i$.
\begin{table*}
\centering \caption{Parameters of TF regressions (5) for the 2MASS
magnitudes and the dipole characteristics for 921 flat galaxies.}
\begin{tabular}{|l|l|l|l|l|l|l|l|} \hline

&\multicolumn{1}{|c|}{$\sigma_{TF}$}& \multicolumn{1}{|c|}{$C_2$}&
\multicolumn{1}{|c|}{$\sigma_{V}$}& \multicolumn{1}{|c|}{$V$}&
\multicolumn{1}{|c|}{$l$}& \multicolumn{1}{|c|}{$b$}&
\multicolumn{1}{|c|}{$F$}\\ \hline $J_{fe}$& 0.432& $-7.64\pm$0.14
(3033)& 1023 & $226\pm$62& $295\pm16$& $-2\pm$13& 4.5 \\
$J_{20fe}$& 0.466 & $-8.20\pm0.15$ (2993)& 1111 & $271\pm$67&
$301\pm$15& $-5\pm12$ & 5.4\\ $J_{ext}$& 0.473 & $-7.83\pm0.15$
(2657)& 1113 & $254\pm$66& $299\pm$16& $+6\pm13$ & 5.0\\ \hline
 $H_{fe}$& 0.446& $-7.84\pm0.14$
(2995)& 1045 & $242\pm64$& $289\pm15 $& $-12\pm14$& 4.7\\
$H_{20fe}$& 0.481& $-8.42\pm$0.15 (2978)& 1129 & $263\pm69 $&
$298\pm15 $& $-9\pm14$& 4.9 \\ $H_{ext}$& 0.495& $-8.11\pm$0.16
(2601)& 1156 & $243\pm71 $& $289\pm17 $& $-4\pm14$& 4.0 \\ \hline
 $K_{fe}$& 0.453 & $-8.10\pm$0.15
(3101)& 1067& $195\pm$65 & $288\pm$19& $-4\pm$16& 3.0 \\
$K_{20fe}$& 0.491& $-8.68\pm$0.16 (3033)  & 1151 & $240\pm70$&
$295\pm$17 & $-5\pm15 $& 3.9 \\ $K_{ext}$& 0.487& $-8.28\pm$0.16
(2802)  & 1141 & $183\pm68$& $295\pm$21 & $+6\pm18 $& 2.5 \\
 \hline
\end{tabular}
\end{table*}

Comparing the data in Tables 3 - 5 allows us to conclude:

1) The cleaning of the sample diminishes significantly the
dispersion $\sigma_{TF}$ ($\sim $ 2 times) and the bulk motion
velocity modulus (about 1.5 -- 2.5 times).

2) In all procedures (the cleaning of the sample, the use of
regression with other magnitudes) the galactic longitude of apex
changes little, within a 20$\degr$ range.

3) The cleaning of the sample  moves the apex to the galactic
equator.

\section{Peculiar velocity field}

Basing on four-parametric regression (5) and derived peculiar
velocities, we built the peculiar velocity field of the RFGC
galaxies for the sample $N$=921.
\begin{figure*}
\centering
\includegraphics[width=10.5cm]{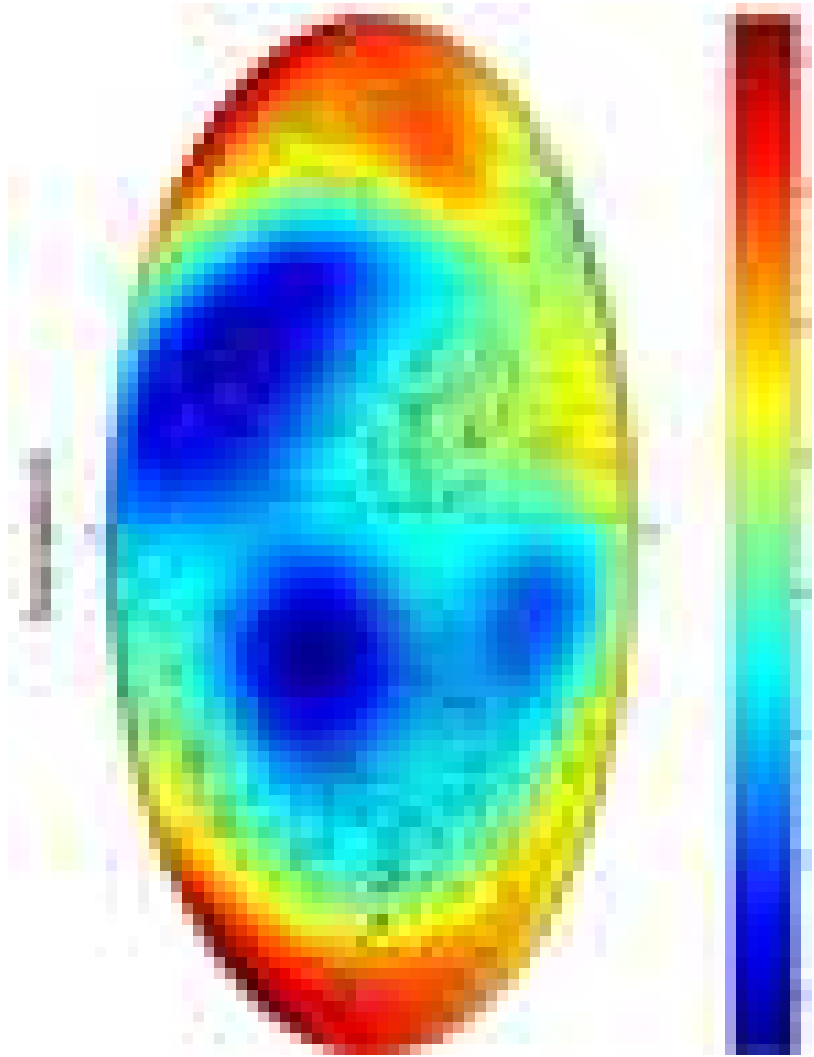}
\vspace{5mm}
\caption{ A smoothed peculiar velocity field for 921 RFGC galaxies.
The galaxies with $V_{pec}>0$ and $V_{pec}<0$ are designated as open
and filled circles, respectively. The "Zone of Avoidance" is indicated
by a solid line.}
\end{figure*}
In Fig.4 we present a smoothed peculiar velocity field for them
in supergalactic coordinates SGL, SGB. The solid line indicates the
"Zone of Avoidance" within $\pm$ 10 degrees around the Galactic equator,
where effects from reddening, stellar contamination and incompleteness
are significant. The galaxies with
$V_{pec}>0$ are designated by open circles, and the galaxies
with $V_{pec }< 0$ with filled circles. The smoothing was done
with a Gaussian filter of 20$\degr$  window. The color scale at
the bottom Fig.4 shows a range of average peculiar velocities from
$-$200 km s$^{-1}$ to +500 km s$^{-1}$.

\begin{figure*}
\centering
\includegraphics[width=8.5cm]{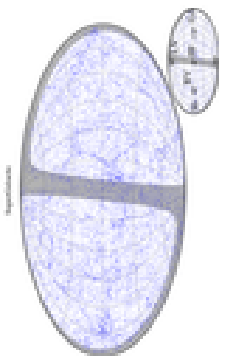}
\caption{
The sky distribution in $SGL, SGB$ of 22361
galaxies from 2MASS survey with an axial ratio $sba<0.3$. Right
bottom the positions of the Great Attractor (GA), the
Pisces-Perseus (PP), the Shapley concentration (Sh), as well as
the Local Void (LV), the Bootes Void (BV) are marked. The "Zone of Avoidance"
within Galactic latitude of $\pm$ 10 degrees is shown by grey color.}
\end{figure*}

In Fig.5 we give the sky distribution in SGL, SGB of 22361
galaxies from the 2MASS survey selected by infrared axial ratio criterion
$sba<0.3$. Bottom right the positions of the Great Attractor (GA),
the Pisces-Perseus supercluster (PP), the Shapley concentration
(Sh), as well as the Local Void (LV) and the Bootes Void (BV) are
marked.
Comparison of Figs. 4 and 5 shows that the regions with positive
and negative smoothed velocities reveal, as a whole, the galaxy
overdensed and underdensed regions. The regions with maximal
positive $V_{pec}$ lie in the direction of the GA and the Shapely
Concentration. We should draw a cautionary note that the obscuring
effects on the Milky Way are also at its maximum towards these
regions (e.g., Abell 3627 of the GA region is well within the "Zone
of Avoidance"), rendering large regions of the sky incomplete in the
2MASS XSC and the RFGC. Significant negative $V_{pec}$ are seen towards
the voids, and low positive velocities ($\sim $100 -- 150 km s$^{-1
}$) are seen in the PP and Hercules - Coma - Corona Borealis
superclusters.

\section{Comparison with other data}

To estimate the amount and quality of the observational material
by which one was used to  determine the bulk motion parameters, we
introduce the value G, meaning the sample goodness

$$G = (N/ 100 )^{1/2} / \sigma_{TF}. \eqno(6)$$

Here $N$ is the number of galaxies under study and $\sigma_{TF }$
is their dispersion on the TF diagram.  From the papers of
different authors, we took the sample of spiral galaxies lying
mainly in sparse regions.  In columns of Table 6 we give: (1) --
designation of the sample, (2) -- color band used, (3) -- number
of galaxies in the sample, (4) -- the dispersion on the TF
diagram, (5) -- the sample goodness, (6) -- source of data. The
upper four lines note the samples included in the Mark III catalog
with their respective name. The data for them are taken from the
Tables 1 and 3 in Willick et al. (1997). The next three samples
are interesting because of their relativelly large depth ($cz \sim
$ 10000 km s$^{-1 }$). The sample of Bamford (2002) was selected
from the SCI sample of Giovanelli et al. (1997b). From 782 S
galaxies placed in the clusters or groups Bamford selected 153
galaxies with available 2MASS photometry. Besides Bamford's data,
we have not yet encountered published TF results using 2MASS.

\begin{table*}
\centering
\caption{Values of $G$ for several samples of spiral galaxies}
\begin{tabular}{|l|l|l|l|l|l|} \hline
Sample & Band & $N$  & $\sigma_{TF}$& $G$& Reference \\ \hline A82
field sample & $H_{-0.5}$& 359  & 0.47 & 4.0  & Willick et
al.,1997 (Mark III)
 \\                    MAT field
sample  & $I$                 & 1355              & 0.43 & 8.6 &
Mark III
\\ W91PP field sample & $r$                  & 326
& 0.38               & 5.0                & Mark III           \\
CF field sample    & $r$                  & 321                &
0.38               & 4.7                & Mark III           \\
Great Wall spirals & $I$                  & 172                &
0.32               & 4.2                & Dell'Antonio et al, 1996
\\ Spirals in clusters & $I$ & 522 & 0.38 & 6.0 & Dale et al, 1999
\\ Spirals in clusters     & $J, H, K_s$
& 153                  & 0.48                 & 2.6
& Bamford, 2002        \\ RFGC                & $J, H, K_s$
& 971                  & 0.42                 & 7.4
& This work            \\ \hline
\end{tabular}
\end{table*}
In last line we give the results obtained in this paper. As seen,
the data goodness varies  over a rather wide range, and the
enormous samples have, naturally, the higher value of $G$.

Note, that different authors use various manners for creating
their ``pure'' samples. In case of our sample, we have not yet
made a detailed analysis of the observable errors, nor reduced the
HI line measurements to a common system, nor analyzed the
Malmquist bias. We excluded the outlying galaxies only
statistically  because of their greater than 3$\sigma$ deviations
from the TF fit and large calculated $V_{pec}$. Nevertheless, our
data are of high goodness ($G$=7.4 in comparison with the median
4.8 for the samples in Table 6). As the minimal scatter on the TF
diagram has a finite intrinsic value (0.25 -- 0.30 mag), the improvement
of the data goodness is undoubtedly associated with the significantly
larger samples.

Let us compare briefly our results with the recent literature data.
The most representative sample in Table 6, the Mark III catalog, yields
the bulk motion $V = 370\pm110$ km s$^{-1 }$ towards $l = 305 \degr$,
$b = +14 \degr$ in the volume of $Hr <5000$ km s$^{-1 }$
(Dekel et al.,1999). Our dipole solution is consistent with the
Mark III data.
 Zaroubi (2002) presents a review of the latest results of
galaxy bulk motion measurements accomplished by different authors
on various observational data. We compare our result with other
dipole determinations using Fig.1 and Table 1 from Zaroubi (2002).
Our RFGC sample can be considered as relatively distant (its depth
reaches 18000 km s$^{-1}$). At the comparable distances ($\sim
$100 -- 150 Mpc h$^{ - 1})$ according to Zaroubi compilation, the
bulk motion parameters were determined for the samples of distant
elliptical galaxies, EFAR (Colless et al., 2001), distant Abell
clusters, LP10 (Willick, 1999), point sources from IRAS $z$-catalog
(PSCz)(Saunders et al. 2000, Branchini et al. 2000), spiral galaxies in
clusters and superclusters (Dale et al., 1999), and distant clusters,
LP (Lauer \& Postman, 1994).

The low amplitude of bulk motion, $V  = 199\pm 62$ km s$^{-1}$,
determined in the present paper for 971 RFGC galaxies with
$Hr<$18000 km s$^{-1}$ and $V_{3K}<$ 18000 km s$^{-1}$, is in
agreement with the results  $V \sim $(0 -- 200) km s$^{-1}$
obtained for the samples EFAR, PSCz, SCI/SCII, as well as SNIa
(Riess et al., 1997) which is consistent with the assumption
of the flow field convergence to the CMB rest-frame at 100 h$^{ -
1}$ Mpc. However, our result differs strongly from the bulk
velocity obtained from LP10 and LP samples ($V \sim $700 km
s$^{-1}$). The low bulk velocity at large scales has been
predicted by popular theories of structure formation in the
cosmological model with cold dark matter and cosmological
constant.

The bulk velocity apex obtained in this paper, $l =
301\degr\pm18\degr$, $b = -2\degr\pm15\degr$, is located near the
apex position ($l = 282\degr$, $b = -8\degr$) derived from SNIa by
Riess et al. (1997) and, within the errors, lies near GA and the
massive cluster Abell 3627.

\section{Conclusion}

We used the 2MASS TF relations to obtain parameters of bulk motion
for flat edge-on galaxies from RFGC. Because about 71{\%} of the
all RFGC sample have $J$, $H$, $K_{s}$  magnitudes from 2MASS,
the size of our sample under study is restricted only by the available
HI line width data. At present, in there is a sample of 1141 all-sky
distributed RFGC galaxies in our disposal with both 2MASS
magnitudes and velocity/line-width estimates.

After excluding about 15{\%} of the sample (dwarf galaxies, very
distant ones, and also the objects which have deviation over
3$\sigma$ on the TF diagram), we built a set of multi- parametric
TF relations using the Kron, isophotal, and extended $J$, $H$,
$K_{s}$ magnitudes. The minimal dispersion on the TF diagram is
shown to be that for Kron $J_{fe}$ magnitude ($\sigma_{TF}$ =
0.422$^{m})$.

For the RFGC sample of $N$=971 galaxies with $V_{3K}<18000$ km
s$^{-1}$, the bulk velocity and apex position are: $ V $ =
199$\pm$ 61 km s$^{-1}$, $l = 301\degr\pm 18\degr$,
 $b =-2\degr\pm 15\degr$. The parameters of bulk motion change
  insignificantly
with the use of other 2MASS magnitudes $J_{20fe}$, $H_{20fe}$,
$K_{20fe}$, $H_{fe}$, $K_{fe}$, $J_{ext}$, $H_{ext}$, $K_{ext}$
($V $ changes within $\pm$ 50 km s$^{-1}$, $l$ and $b$ within
$\pm15\degr)$. This result remains robust to decreasing the
sample depth. In particular, for 921 RFGC galaxies with $V_{3K} <
$12000 km/s we obtained $ V $ = 226$\pm 62$ km s$^{-1}$, $l =
295\degr\pm 16\degr$, $b = -2\degr\pm 13\degr$. Within the errors,
our estimates of the bulk velocity are in agreement with the data
for the EFAR, PSCz, SCI/SCII samples. Besides, the apex position
and $ V $ are consistent with the results obtained for SNIa.

The two-dimensional smoothed peculiar velocity field well traces
the large-scale density variations in the galaxy distribution,
e.g. Great Attractor, Pisces-Perseus an Hercules-Coma-Corona
Borealis superclusters, Shapley concentration, Bootes and Local
voids.

Thus, we show by the example of the RFGC catalogue that the 2MASS
Tully-Fisher relation can be used successfully in studying galaxy
cosmic flows. The next obvious step is to complete the
observations of radial velocities and HI line widths for the remaining
RFGC galaxies.

\acknowledgements {We thank Dmitry Makarov for help in the Fig. 4
and Fig.5 design.

This paper makes use of data products from the Two Micron All Sky
Survey, which is a joint project of the University of
Massachusetts and the Infrared Processing and Analysis
Center/California Institute of Technology, funded by the National
Aeronautics and Space Administration  and the National Science
Foundation.

We have made use of the LEDA database (\underline
{http://leda.univ-lyon1.fr}).

This research was partially supported by DFG-RFBR grant 436RUS
113/701/0-1.}

{}

\end{document}